\documentclass[twocolumn,tighten]{aastex63}
\usepackage{amsmath}
\usepackage{threeparttable}
\usepackage{float}
\usepackage{graphicx}
\usepackage{CJK}

\newcommand{\package}[1]{\tt{#1}}

\newcommand{\MgI}{\hbox{{\rm Mg}\kern 0.1em{\sc i}}}

\defcitealias{Paper2}{Matthee \& Naidu et al.}

\shorttitle{Disrupted Dwarfs: Mass-Metallicity-Alpha}
\shortauthors{Naidu et al.}

\begin{document}
\begin{CJK*}{UTF8}{gbsn}

\title{Live Fast, Die $\alpha$-Enhanced: The Mass-Metallicity-$\alpha$ Relation of the Milky Way's Disrupted Dwarf Galaxies}

\correspondingauthor{Rohan P. Naidu}
\email{rohan.naidu@cfa.harvard.edu}
\author[0000-0003-3997-5705]{Rohan P. Naidu}
\affiliation{Center for Astrophysics $|$ Harvard \& Smithsonian, 60 Garden Street, Cambridge, MA 02138, USA}
\author[0000-0002-1590-8551]{Charlie Conroy}
\affiliation{Center for Astrophysics $|$ Harvard \& Smithsonian, 60 Garden Street, Cambridge, MA 02138, USA}
\author[0000-0002-7846-9787]{Ana Bonaca}
\affiliation{Observatories of the Carnegie Institution for Science, 813 Santa Barbara St., Pasadena, CA 91101, USA}
\author[0000-0002-5177-727X]{Dennis Zaritsky}
\affiliation{Steward Observatory, University of Arizona, 933 North Cherry Avenue, Tucson, AZ 85721-0065, USA}
\author[0000-0001-5082-9536]{Yuan-Sen Ting (丁源森)}
\affiliation{Research School of Astronomy \& Astrophysics, Australian National University, Cotter Road, Weston Creek, ACT 2611, Canberra, Australia}
\affiliation{Research School of Computer Science, Australian National University, Acton ACT 2601, Australia}
\author[0000-0003-2352-3202]{Nelson Caldwell}
\affiliation{Center for Astrophysics $|$ Harvard \& Smithsonian, 60 Garden Street, Cambridge, MA 02138, USA}
\author[0000-0002-1617-8917]{Phillip Cargile}
\affiliation{Center for Astrophysics $|$ Harvard \& Smithsonian, 60 Garden Street, Cambridge, MA 02138, USA}
\author[0000-0003-2573-9832]{Joshua S. Speagle (\begin{CJK*}{UTF8}{gbsn}沈佳士\ignorespacesafterend\end{CJK*})}
\altaffiliation{Banting \& Dunlap Fellow}
\affiliation{Department of Statistical Sciences, University of Toronto, Toronto, ON M5S 3G3, Canada}
\affiliation{David A. Dunlap Department of Astronomy \& Astrophysics, University of Toronto, Toronto, ON M5S 3H4, Canada}
\affiliation{Dunlap Institute for Astronomy \& Astrophysics, University of Toronto, Toronto, ON M5S 3H4, Canada}
\author[0000-0002-0572-8012]{Vedant Chandra}
\affiliation{Center for Astrophysics $|$ Harvard \& Smithsonian, 60 Garden Street, Cambridge, MA 02138, USA}
\author[0000-0002-9280-7594]{Benjamin D. Johnson}
\affiliation{Center for Astrophysics $|$ Harvard \& Smithsonian, 60 Garden Street, Cambridge, MA 02138, USA}
\author[0000-0002-0721-6715]{Turner Woody}
\affiliation{Center for Astrophysics $|$ Harvard \& Smithsonian, 60 Garden Street, Cambridge, MA 02138, USA}
\author[0000-0002-6800-5778]{Jiwon Jesse Han}
\affiliation{Center for Astrophysics $|$ Harvard \& Smithsonian, 60 Garden Street, Cambridge, MA 02138, USA}

\begin{abstract}
The Milky Way's satellite galaxies (``surviving dwarfs") have been studied for decades as unique probes of chemical evolution in the low-mass regime. Here we extend such studies to the ``disrupted dwarfs", whose debris constitutes the stellar halo. We present abundances ([Fe/H], [$\alpha$/Fe]) and stellar masses for nine disrupted dwarfs with $M_{\rm{\star}}\approx10^{6}-10^{9}M_{\rm{\odot}}$ from the H3 Survey (Sagittarius, \textit{Gaia}-Sausage-Enceladus, Helmi Streams, Sequoia, Wukong/LMS-1, Cetus, Thamnos, I'itoi, Orphan/Chenab). The surviving and disrupted dwarfs are chemically distinct: at fixed mass, the disrupted dwarfs are systematically metal-poor and $\alpha$-enhanced. The disrupted dwarfs define a mass-metallicity relation (MZR) with a similar slope as the $z=0$ MZR followed by the surviving dwarfs, but offset to lower metallicities by $\Delta$[Fe/H]${\approx}0.3{-}0.4$ dex. Dwarfs with larger offsets from the $z=0$ MZR are more $\alpha$-enhanced with [$\alpha$/Fe] = $0.43^{+0.09}_{-0.09}\times\Delta\rm{[Fe/H]}+0.08^{+0.03}_{-0.03}$. In simulations as well as observations, galaxies with higher $\Delta$[Fe/H] formed at higher redshifts -- exploiting this, we infer the disrupted dwarfs have typical star-formation truncation redshifts of $z_{\rm{trunc}}{\sim}1-2$. We compare the chemically inferred $z_{\rm{trunc}}$ with dynamically inferred accretion redshifts and find almost all dwarfs are quenched only after accretion. The differences between disrupted and surviving dwarfs are likely because the disrupted dwarfs assembled their mass rapidly, at higher redshifts, and within denser dark matter halos that formed closer to the Galaxy. Our results place novel archaeological constraints on low-mass galaxies inaccessible to direct high-$z$ studies: (i) the redshift evolution of the MZR along parallel tracks but offset to lower metallicities extends to $M_{\rm{\star}}\approx10^{6}-10^{9}M_{\rm{\odot}}$; (ii) galaxies at $z\approx2-3$ are $\alpha$-enhanced with [$\alpha$/Fe]$\approx$0.4.
\end{abstract}


\section{Introduction}
\label{sec:introduction}

In the $\Lambda$CDM paradigm, galaxies like the Milky Way (MW) grow by assimilating smaller galaxies \citep[e.g.,][]{White91}. While some of these accreted galaxies orbit the MW largely intact (``surviving dwarfs"), the vast majority ($\gtrsim70-80\%$, e.g., \citealt[][]{Fattahi20,Santistevan20}) are predicted to have been tidally shredded (``disrupted dwarfs"). The remains of these disrupted dwarfs are expected to comprise the bulk of the stellar halo \citep[e.g.,][]{bj05_1,Cooper10,Monachesi19,Font20}.

A puzzle in this context was revealed by chemical abundance surveys of the surviving dwarfs. These dwarfs were found to be deficient in $\alpha$-elements, often with sub-solar [$\alpha$/Fe] compared to halo stars with [$\alpha$/Fe]$\approx0.2-0.3$ at similar metallicity \citep[e.g.,][]{Fulbright02,Shetrone03,Tolstoy03,Venn04,Geisler05}. If the halo was built out of dwarf accretion, why is it chemically distinct from the surviving dwarfs? A proposed solution came from simulations that predicted that the majority of halo stars arose from a handful of relatively massive, $\alpha$-enhanced galaxies accreted at $z\gtrsim2$ \citep[e.g.,][]{Font06}. For instance, \citet[][]{Robertson05} argued that the observed halo abundance pattern was likely driven by a handful of massive $M_{\rm{\star}}\approx2.5\times10^{8}M_{\rm{\odot}}$ dwarfs disrupted $\approx10$ Gyrs ago. The predicted $\alpha$-enhancement was due to their rapid star-formation histories necessitated by their early accretion times. In short, these solutions posited that the disrupted dwarfs had distinct stellar populations from the surviving dwarfs.

Tracing every star in the halo back to its parent dwarf galaxy has been a long-harbored ambition of Galactic astronomy \citep[e.g.,][]{Johnston96,Helmi00,Brown05}. Thanks to \textit{Gaia} and complementary spectroscopic surveys this ambition is now being fulfilled -- the majority of halo stars have been associated with distinct parent galaxies \citep[e.g.,][]{Naidu20,Yuan20b,Bonaca21,An21,Horta21,Ji21,Malhan22}. For instance, one of the massive dwarfs anticipated by \citet[][]{Robertson05} is now known to be \textit{Gaia}-Sausage Enceladus (GSE, \citealt[][]{Belokurov18,Helmi18,Naidu21}), which merged with the MW at $z\approx1-2$ \citep[e.g.,][]{Gallart19,Bonaca20,Belokurov20,Montalban21,Xiang22}. GSE ([$\alpha$/Fe]$\approx0.2$) comprises $\approx50\%$ of the stellar halo, and the vast majority of the [Fe/H]$<-1$ halo in the solar neighborhood \citep[e.g.,][]{Naidu20}. This single accreted object largely explains the higher [$\alpha$/Fe] of the local halo compared to surviving dwarfs.

With the halo now resolved into a collection of distinct progenitors, we can directly compare stellar populations of disrupted and surviving dwarfs. A large body of theoretical work has predicted significant differences -- for instance, the disrupted dwarfs are expected to have been accreted much earlier, largely at $z\gtrsim1$, whereas the surviving dwarfs are generally expected to be more recent arrivals with extended star-formation histories \citep[e.g.,][]{ Sales07,Tissera12,Fattahi20}. These differences are expected to manifest in chemical abundance patterns \citep[e.g.,][]{Robertson05,Font06}. Indeed, several observational studies have contrasted the chemistry (e.g., $\alpha$-abundances) and make-up (e.g., relative fractions of RR Lyrae subtypes) of halo stars as a whole vs. the surviving dwarfs, finding stark differences \citep[e.g.,][]{Venn04,Tolstoy03,Tolstoy09,Zinn14,Stetson14,Fiorentino15,Fiorentino17,Belokurov18rrl}. Here we disaggregate the halo to offer an apples-to-apples comparison of surviving vs. disrupted dwarfs at fixed stellar mass.

Independent of their relationship to the MW satellites, the disrupted dwarfs provide convenient access to early-Universe stellar abundances at a spatial resolution (star-by-star) and mass-range ($M_{\rm{\star}}\lesssim10^{9} M_{\rm{\odot}}$) currently inaccessible to direct high-$z$ studies. While the surviving dwarfs provide the clearest view of e.g., the $z=0$ stellar mass-stellar metallicity relation (MZR, e.g., \citealt{Kirby13}), the disrupted dwarfs are expected to have formed and disintegrated at higher redshifts on average \citep[e.g.,][]{Fattahi20}. The disrupted dwarfs are already being used in ``high-$z$" studies to make timing arguments that exploit the disruption redshift and infer that both core-collapse supernovae (CCSNe) and neutron star mergers produce $r$-process elements \citep[e.g.,][]{Matsuno21,Naidu22}. Here we derive novel archaeological constraints on the stellar chemistry of low-mass galaxies at $z\approx1-3$ complementary to direct high-$z$ studies.

This paper is organized as follows. In \S\ref{sec:data} we assemble [Fe/H], [$\alpha$/Fe], and stellar masses for the surviving and disrupted dwarfs. In \S\ref{sec:results} we describe the four interlinked figures in this paper, and then interpret them in \S\ref{sec:discussion} -- \S\ref{sec:whymetalpoor} and \S\ref{sec:butwhymetalpoor} explain the differences between the disrupted and surviving dwarfs, \S\ref{sec:sequencing} compares accretion and quenching redshifts, and \S\ref{sec:highzarchaeology} describes the archaeological connection to high-$z$ stellar abundances. ``Dwarfs" always denotes dwarf galaxies, and never refers to low luminosity stars. ``Z" denotes metallicity, and ``$z$" stands for redshift. We use medians to describe the central values of distributions, with uncertainties on the median (16$^{\rm{th}}$ and 84$^{\rm{th}}$ percentiles) from bootstrap resampling. In converting between redshifts and lookback times we adopt the \citet[][]{Planck18} cosmology.

\begin{deluxetable*}{lrrrrrrrrrr}
\label{table:summary}
\tabletypesize{\footnotesize}
\tablecaption{Properties of disrupted dwarf galaxies from the H3 Survey.}
\tablehead{
\colhead{Galaxy} & \colhead{$\log(M_{\star}/M_{\rm{\odot}})$} & \colhead{[Fe/H]} & \colhead{[$\alpha$/Fe]} & \colhead{$\Delta$[Fe/H]} & \colhead{$z_{\rm{trunc}}$} & \colhead{$z_{\rm{acc}}$} & 
\colhead{$N_{\rm{\star}}$}}
\startdata
\vspace{-0.2cm}  \\
Sagittarius & 8.8 & -0.96$^{+0.03}_{-0.03}$ & 0.12$^{+0.00}_{-0.01}$ & 0.01 & $<0.1$ & 0.6 & 675\\
\textit{Gaia}-Sausage-Enceladus & 8.7 & -1.18$^{+0.00}_{-0.01}$ & 0.21$^{+0.00}_{-0.00}$ & 0.30 & 1.2 & 2.0 & 2684\\
Helmi Streams & 8.0 & -1.28$^{+0.02}_{-0.04}$ & 0.15$^{+0.01}_{-0.01}$ & 0.19 & 0.8 & 1.0 & 91\\
Sequoia & 7.2 & -1.59$^{+0.03}_{-0.04}$ & 0.14$^{+0.03}_{-0.01}$ & 0.25 & 0.7 & -- & 72\\
Wukong/LMS-1 & 7.1 & -1.58$^{+0.06}_{-0.03}$ & 0.24$^{+0.03}_{-0.01}$ & 0.23 & 0.9 & 1.1 & 111\\
Cetus & 7.0 & -1.85$^{+0.04}_{-0.04}$ & 0.29$^{+0.01}_{-0.02}$ & 0.45 & 2.3 & 0.5 & 56\\
Thamnos & 6.7 & -1.90$^{+0.08}_{-0.06}$ & 0.29$^{+0.04}_{-0.01}$ & 0.41 & 1.9 & $\gtrsim2$ & 32\\
I'itoi & 6.3 & -2.39$^{+0.07}_{-0.07}$ & 0.38$^{+0.04}_{-0.01}$ & 0.79 & $>6$ & -- & 65\\
Orphan/Chenab & 6.1 & -1.75$^{+0.10}_{-0.20}$ & 0.04$^{+0.10}_{-0.07}$ & 0.10 & $0.3$ & -- & 11
\enddata
\tablecomments{Stellar masses for Sgr, GSE, and Helmi Streams are from tailored N-body simulations, and for the remaining galaxies from star-counts (see \S\ref{sec:datadisrupted}). $\Delta$[Fe/H] measures the vertical offset of a galaxy from the $z=0$ MZR in \citet[][]{Kirby13}, with metal-poor (metal-rich) galaxies defined to have positive (negative) $\Delta$[Fe/H] -- see Eqn. \ref{eqn:deltafeh}. $z_{\rm{trunc}}$ is the SFH truncation redshift, defined as the redshift at which the galaxy lies on the MZR from the FIRE simulations \citep[][]{Ma16MZR}, see Eqn. \ref{eqn:ztrunc}. The accretion redshift ($z_{\rm{acc}}$) is when the disrupted system makes its first pericentric passage, and is based on N-body simulations listed in \S\ref{sec:datadisrupted}. $N_{\rm{\star}}$ is the number of stars identified in \citet[][]{Naidu20} as part of each structure, except for Cetus and Orphan/Chenab that are selected from the current H3 giants sample (see \S\ref{sec:datadisrupted}).}
\end{deluxetable*}

\begin{deluxetable*}{lrrrrrrrrrr}
\label{table:surviving}
\tabletypesize{\footnotesize}
\tablecaption{Properties of surviving dwarf galaxies.}
\tablehead{
\colhead{Galaxy} & \colhead{$\log(M_{\star}/M_{\rm{\odot}})$} & \colhead{[Fe/H]} & \colhead{[$\alpha$/Fe]} & \colhead{$\Delta$[Fe/H]} & \colhead{$z_{\rm{trunc}}$} & \colhead{$z_{\rm{acc}}$} & 
\colhead{$N_{\rm{Fe}}$} & \colhead{$N_{\rm{Mg}}$}} 
\startdata
\vspace{-0.2cm}
Large Magellanic Cloud & 9.4 & -0.71$^{+0.00}_{-0.01}$ & 0.04$^{+0.00}_{-0.00}$ & 0.05 & 0.15 & 0.11 & 3908 & 3907\vspace{0.2cm}\\
Small Magellanic Cloud & 8.5 & -1.08$^{+0.00}_{-0.01}$ & 0.0$^{+0.00}_{-0.00}$ & 0.14 & 0.46 & 0.11 & 1143 & 1142\\
Fornax & 7.4 & -1.04$^{+0.01}_{-0.01}$ & -0.13$^{+0.01}_{-0.01}$ & -0.23 & -- & 2.12 & 672 & 372\\
Leo I & 6.7 & -1.45$^{+0.01}_{-0.01}$ & 0.16$^{+0.03}_{-0.04}$ & -0.03 & -- & 0.18 & 814 & 170\\
Sculptor & 6.6 & -1.68$^{+0.01}_{-0.01}$ & 0.20$^{+0.02}_{-0.05}$ & 0.17 & 0.57 & 2.05 & 375 & 96\\
Antlia 2 & 6.2 & -1.90$^{+0.04}_{-0.04}$ & -- & 0.28 & 1.06 & -- & 283 & --\\
Leo II & 6.1 & -1.63$^{+0.01}_{-0.01}$ & 0.14$^{+0.03}_{-0.07}$ & -0.04 & -- & 0.68 & 256 & 54\\
Carina & 6.0 & -1.72$^{+0.01}_{-0.01}$ & 0.18$^{+0.03}_{-0.03}$ & 0.04 & 0.13 & 1.66 & 437 & 60\\
Sextans & 5.8 & -1.94$^{+0.01}_{-0.01}$ & 0.11$^{+0.06}_{-0.03}$ & 0.2 & 0.72 & -- & 123 & 65\\
Ursa Minor & 5.7 & -2.13$^{+0.01}_{-0.01}$ & 0.32$^{+0.07}_{-0.01}$ & 0.36 & 1.53 & 1.81 & 190 & 26\\
Crater 2 & 5.6 & -2.16$^{+0.04}_{-0.04}$ & -- & 0.34 & 1.39 & -- & 141 & --\\
Draco & 5.5 & -1.98$^{+0.01}_{-0.01}$ & 0.12$^{+0.07}_{-0.07}$ & 0.14 & 0.48 & 1.34 & 269 & 14\\
Canes Venatici I & 5.5 & -1.91$^{+0.01}_{-0.01}$ & -- & 0.06 & 0.2 & 1.09 & 151 & --\\
\enddata
\tablecomments{See caption of Table \ref{table:summary} for column definitions. See \S\ref{sec:surviving} for references for $\log(M_{\star}/M_{\rm{\odot}})$, [Fe/H], and [Mg/Fe]. $z_{\rm{acc}}$ values are from \citet[][]{Fillingham19}. $N_{\rm{Fe}}$ and $N_{\rm{Mg}}$ are the number of stars used to estimate [Fe/H] and [$\alpha$/Fe] respectively. A negative $\Delta$[Fe/H] means the galaxy is more metal-rich than expected from the $z=0$ MZR -- the corresponding $z_{\rm{trunc}}$ from Eqn. \ref{eqn:ztrunc} is undefined, consistent with relatively recent star-formation observed in such  systems (e.g., Fornax).}
\end{deluxetable*}

\section{Data \& Methods}
\label{sec:data}

\subsection{Disrupted galaxy sample}
\label{sec:datadisrupted}
This study builds on results from the H3 Survey \citep{Conroy19} -- an ongoing high-latitude ($|b|>20^{\circ}$), high-resolution ($R=32,000$) spectroscopic survey of $\approx300,000$ stars in the distant ($d_{\rm{helio}}\approx2-100$ kpc) Galaxy. H3 is measuring radial velocities precise to $\lesssim$1 km $\rm{s^{-1}}$, [Fe/H] and [$\alpha$/Fe] abundances precise to $\lesssim$0.1 dex, and spectrophotometric distances precise to $\lesssim$10$\%$ \citep{Cargile20}. Combined with \textit{Gaia} proper motions (SNR$>$3 for $>$90$\%$ of the sample), H3 provides full 6D phase-space and 2D chemical-space for detecting and characterizing substructure. \citet[][]{Naidu20} used the H3 sample of giant stars (N=5684, $|b|{>}40^{\circ}$, $d_{\rm{helio}}=3-50$ kpc) to assign almost the entire distant Galaxy to individual structures including known accreted dwarfs (e.g., Sagittarius), as well as hitherto unknown structures (e.g., Wukong).

The sample of disrupted dwarf galaxies analyzed in this work is as defined in \citet[][]{Naidu20}, and includes Sagittarius \citep[][]{Ibata94,LM10, Johnson20}, \textit{Gaia}-Sausage-Enceladus \citep[][]{Belokurov18, Haywood18, Myeong18, Helmi18, Feuillet21,Buder22}, the Helmi Streams \citep[][]{Helmi99, Koppelman19HS, Limberg21}, Sequoia \citep[][]{Myeong19, Matsuno19, Matsuno21Seq, Monty20, Aguado21}, Wukong (independently discovered as LMS-1 in \citealt[][]{Yuan20b}, and studied in \citealt[][]{Malhan21,Malhan22,Shank22}), Thamnos \citep[][]{Koppelman19, Lovdal22, Ruiz-Lara22}, and I'itoi. 

We note three relevant developments since \citet[][]{Naidu20}. The Arjuna structure identified in that work is treated here as the highly retrograde debris of GSE, and is thus included as a part of GSE (see \citealt{Naidu21} for details). With the acquisition of new southern fields covering the dynamically cold Cetus stream \citep[e.g.,][]{Newberg09,Thomas21, Yuan21}, the current H3 giants sample ($>2.5\times$ larger than the sample studied in \citealt[][]{Naidu20}) now has $N=56$ confident Cetus members. Due to its dynamical coherence, Cetus is easily selected as follows:
\begin{align}
    (L_{\rm{y}}>0.7) \land (L_{\rm{x}}>2.0) \land (-3.4<L_{\rm{z}}<-1.4)\nonumber\\\land\ (-1<E_{\rm{tot}}<-0.6), 
\end{align}
where total orbital energy ($E_{\rm{tot}}$) is in units of $10^{5}\ \rm{km^{2}\ s^{-2}}$ and angular momenta ($L_{\rm{x}}$, $L_{\rm{y}}$, $L_{\rm{z}}$) are in units of $10^{3}\ \rm{kpc}\ \rm{km\ s^{-1}}$. Similarly, we now also have $N=11$ confident members in the Orphan/Chenab galaxy \citep[e.g.,][]{Grillmair06orph,Newberg10,Shipp18,Koposov19,Erkal19} that we select as follows:
\begin{align}
(L_{\rm{z}}>-6) \land (L_{\rm{x}}<-3.0) \land (-6.0<L_{\rm{z}}<-3.5)\nonumber\\
\land\ (-0.7<E_{\rm{tot}}<-0.4).    
\end{align}
The median [Fe/H] we derive ($-1.75^{+0.10}_{-0.20}$) is in excellent agreement with the Orphan ($-1.85^{+0.07}_{-0.07}$) and Chenab ($-1.78^{+0.04}_{-0.04}$) samples studied by the S5 Survey using the CaT feature with $R\approx10,000$ spectroscopy \citep{Li19S5,Li22}.

We adopt the [Fe/H] and [$\alpha$/Fe] abundances reported in Table 1 of \citet[][]{Naidu20} based on $\approx10$ stars in the least sampled system (Orphan/Chenab) and $\approx2700$ stars in the best sampled system (GSE). Statistical errors on the median [Fe/H] and [$\alpha$/Fe] are generally small ($<0.05$) except in a handful of cases (see Table \ref{table:summary}).

For total stellar masses we rely on: (i) tailored N-body simulations for the more massive galaxies (GSE, Sgr, Helmi Streams), and (ii) star counts for the other galaxies from \citet[][]{Naidu20}. Tailored N-body simulations have been run for GSE \citep[][]{Naidu21}, Sgr \citep[e.g.,][]{Laporte18}, the Helmi Streams \citep[][]{Koppelman19HS}, Wukong \citep[][]{Malhan21}, Thamnos \citep{Koppelman19}, and Cetus \citep[][]{Chang20}. These simulations qualitatively reproduce the phase-space distribution of debris, but the downside is that their dynamics are largely set by the total mass, and they are thus reliant on stellar mass - halo mass relations (SMHMRs). This is not a problem for the more massive ($M_{\rm{\star}}>10^{8}M_{\rm{\odot}}$) systems (GSE, Sgr, Helmi Streams) -- the SMHMR is relatively well-determined in this regime \citep[e.g.,][]{Behroozi19}, and multiple cross checks on the stellar mass are available (e.g., chemical evolution modeling with hundreds of stars).

For lower mass systems we use relative star counts reported in \citet[][]{Naidu20}. Here we translate their relative star counts to stellar masses assuming GSE has an $M_{\rm{\star}}=5\times10^{8}M_{\rm{\odot}}$. For instance, Wukong's star counts are $\approx2.5\%$ of those of GSE, and so we infer its mass is $\approx1.3\times10^{7}M_{\rm{\odot}}$. We choose GSE as the reference object to derive other systems' masses because its stellar mass has been estimated via multiple independent methods that agree well with each other \citep[e.g.,][]{Helmi18,Kruijssen20,Mackereth20,Naidu21,Callingham22}, and because it is fairly well mixed within the H3 Survey footprint \citep[][]{Han22}. As a sanity check, based on this method, Sgr would have an inferred stellar mass of $\approx3\times10^{8} M_{\rm{\odot}}$, which is in excellent agreement with various literature estimates that span $\approx 2-6 \times10^{8} M_{\rm{\odot}}$ \citep[e.g.,][]{Niederste-Ostholt10,LM10, Kruijssen20, Vasiliev20breath}, and within a factor of 2 of our adopted value ($6\times10^{8} M_{\rm{\odot}}$) from \citealt[][]{Laporte18}. This cross-check inspires confidence that even for relatively unmixed systems our derived stellar masses are reasonable, and gives a sense of the systematic uncertainty for our reported values.

Note that the relative star counts are corrected for the survey selection function (see \S2.3 of \citealt[][]{Naidu20} for details). We also checked the ``orbit bias" across the various low-mass structures is negligible -- i.e., whether the integrated orbits of stars from a particular system over the last 10 Gyrs spend significantly longer/shorter periods within our survey footprint and distance range probed compared to the GSE sample. The exception is the dynamically cold, unmixed Cetus -- our reported stellar mass for Cetus comes with this caveat, i.e., that it may be underestimated by $\approx2\times$. Finally, we correct the relative star-counts for structures selected via sharp cuts on their MDF in \citet[][]{Naidu20} in order to avoid contamination from the low-eccentricity tail of GSE (e.g., I'itoi). We model these MDFs as having multiple components -- i) the GSE MDF model fit in \citet[][]{Naidu20}, and ii) Gaussians centred at the peak metallicity of the dwarfs in question. The only counts that require meaningful revision are I'itoi's ($\approx-30\%$) and Sequoia's ($\approx+35\%$) which translate to modest 0.1 dex shifts in their inferred stellar mass. The resulting stellar masses for all systems are listed in Table \ref{table:summary}.

An important caveat to bear in mind is that the sample studied in \citet[][]{Naidu20} spans $r_{\rm{gal}}\approx6-50$ kpc. That is, we are missing dwarfs entirely contained in the inner few kpc (e.g., Kraken, \citealt[][]{Kruijssen20}) as well as those exclusively inhabiting the outer reaches of the galaxy. Broadly, the systems with the entirety of their debris at $>50$ kpc are likely to be very recent accretion events ($z\lesssim0.5$), and those buried in the Galactic center were likely accreted very early ($z\gtrsim2$) \citep[e.g.,][]{Pfeffer20}. In other words, there is a diverse spectrum of disrupted dwarfs, and our findings must be read as applying to a specific portion of this spectrum corresponding to intermediate accretion redshifts. The mapping between accretion redshift and the distance at which debris is deposited holds to first order, but is scrambled by other factors like the dwarf's orbit, size, and mass (see \citealt[][]{Amorisco17,Naidu21} for parameter studies).

\subsection{Surviving dwarfs sample}
\label{sec:surviving}

The $M_{\rm{\star}}$ range of interest for comparison with the disrupted dwarf sample is $\approx10^{5}-10^{9} M_{\rm{\odot}}$, which includes 13 known Milky Way satellites. For Fornax, Leo I, Leo II, Sculptor, Draco, Sextans, Ursa Minor (UMin), and Canes Venatici I we draw stellar masses (estimated via integrated luminosities) and metallicities from the \citet{Kirby13} compilation that are homogeneously derived from Keck/DEIMOS medium-resolution ($R\approx6500$) spectroscopy. Given that this compilation comprises the majority of our sample, we also adopt the $z=0$ \citet{Kirby13} MZR for comparison. For the remaining five dwarfs of interest not included in \citet{Kirby13}, metallicities and stellar masses are sourced from \citet[][]{vandermarel09,Nidever20,Hasselquist21} for the Magellanic Clouds, \citet[][]{Ji21} for Crater 2 and Antlia 2, and \citet[][]{Koch06,Koch08,deBoer14} for Carina.

We use [Mg/Fe] to approximate [$\alpha$/Fe] since the H3 spectral window (5150 - 5300 \AA) is mostly sensitive to the prominent \MgI\  triplet at 5175 \AA \citep[][]{Cargile20}. We adopt Mg abundances derived from high-resolution spectroscopy when available, combining samples from multiple studies when required to build up numbers. For the Magellanic Clouds we use the sample in \citet{Hasselquist21}, for Fornax we combine \citet[][]{Letarte10,Letarte18,Hendricks14,Hasselquist21}, for Leo I and Leo II we use \citet[][]{Shetrone09} along with the high-SNR subset ($<0.3$ dex errors on [Mg/Fe]) of \citet[][]{Kirby11aFe}, for Sculptor we draw on \citet[][]{Hill19}, for Carina we use \citet[][]{Norris17}, for Sextans we combine \citet[][]{Shetrone03,Tafelmeyer10,Theler20,Mashonkina22}, for UMin we combine \citet{Shetrone03,Sadakane04,Cohen10,Ural15}, and finally for Draco we combine \citet[][]{Shetrone03,Cohen09}. Antlia 2 and  Crater 2 have no [Mg/Fe] reported, and Canes Venatici I has too few stars measured robustly (N=2 in \citealt[][]{Kirby11aFe} with $<0.3$ dex error on [Mg/Fe]) to make a meaningful estimate. Almost all the stars we compile here are giant stars, similar to our disrupted dwarf sample. All values for [Mg/Fe] reported in Table \ref{table:surviving} are medians and associated uncertainties are errors on the median from bootstrap resampling.

\begin{figure*}
\centering
\includegraphics[width=0.75\linewidth]{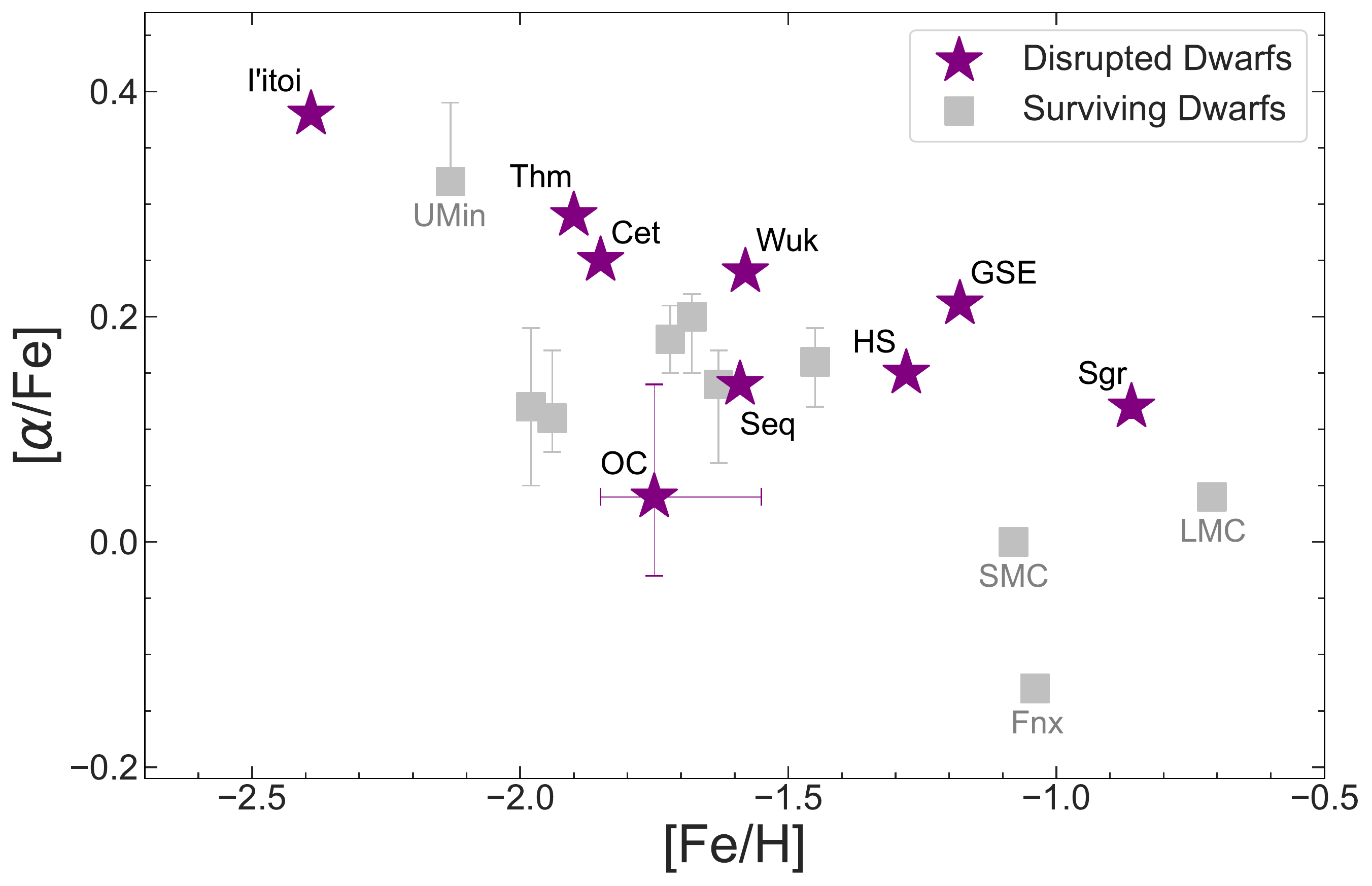}
\caption{[$\alpha$/Fe] vs. [Fe/H] for disrupted dwarfs from the H3 Survey (purple stars) and for surviving dwarfs from the literature (gray squares). At fixed [Fe/H], the disrupted dwarfs are relatively $\alpha$ enhanced compared to the surviving dwarfs, particularly at [Fe/H]$>-1.5$. Note the difference between GSE, which dominates the [Fe/H]$<-1$ halo around the Sun in terms of star counts, and Fornax/SMC that lie at comparable metallicity. This difference explains the well-known result \citep[e.g.,][]{Venn04} that the stellar halo is $\alpha$-enhanced at fixed metallicity compared to stars in the surviving dwarfs.}
\label{fig:classical}
\end{figure*}

\begin{figure*}
\centering
\includegraphics[width=\linewidth]{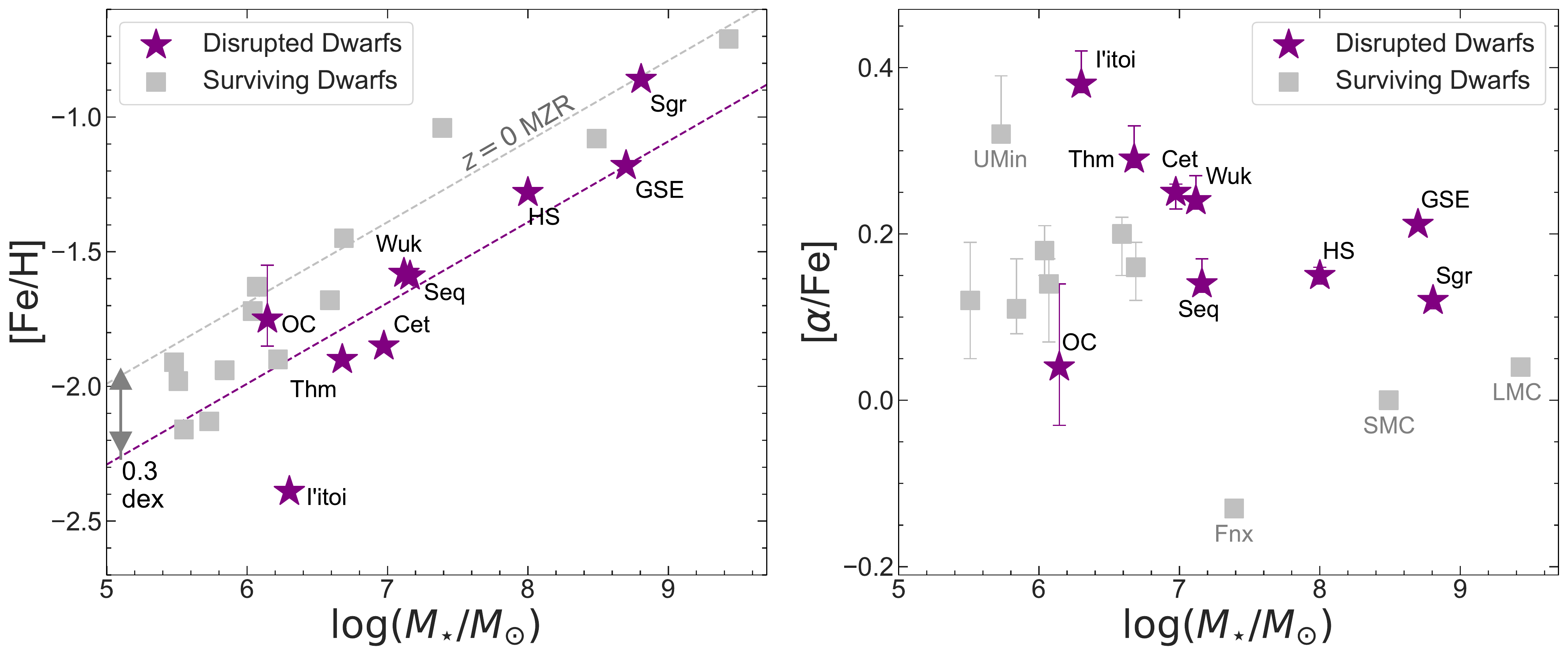}
\caption{\textbf{Left:} Stellar metallicity vs. stellar mass for the surviving (gray) and disrupted (purple) dwarfs. The gray line is the $z=0$ MZR \citep{Kirby13}, while the purple line is offset by 0.3 dex, which is the median offset of the disrupted dwarfs. Sgr and I'itoi are notable outliers, likely due to their low and high disruption redshifts respectively (see Fig. \ref{fig:delta}). \textbf{Right:} [$\alpha$/Fe] vs. stellar mass. At fixed mass, the disrupted dwarfs are systematically $\alpha$-enhanced compared to the surviving dwarfs, particularly at $M_{\rm{\star}}>10^{7} M_{\rm{\odot}}$. UMin and Orphan/Chenab are notable exceptions. Like most of the disrupted dwarfs, UMin had a rapid star-formation history ($\approx400$ Myrs, \citealt[][]{Kirby11aFe}), and was likely accreted very early ($z\approx2$, \citealt[][]{Fillingham19}), while on the other hand, Orphan/Chenab appears to be a relatively recent arrival based on its high orbital energy (apocenter of $\approx70$ kpc), which is comparable to Sgr, and typical for $z<0.5-1$ accretion \citep[e.g.,][]{Pfeffer20}.}
\label{fig:mzr}
\end{figure*}

\section{Results}
\label{sec:results}

In Figure \ref{fig:classical} we plot [Fe/H] vs. [$\alpha$/Fe] for disrupted dwarfs (purple stars) and contrast them with the surviving dwarfs (gray squares). Generally, the disrupted dwarfs are $\alpha$-enhanced compared to the surviving dwarfs. This is particularly apparent at [Fe/H]$\gtrsim-1.4$ with GSE, the Helmi Streams and Sagittarius lying $\approx0.2-0.3$ dex higher than Fornax and the Magellanic Clouds that have comparable metallicities. 

Figure \ref{fig:classical} explains the well-known result \citep[e.g.,][]{Venn04,Tolstoy03,Tolstoy09} that at fixed metallicity the local halo is preferentially $\alpha$-enhanced compared to surviving dwarf stars. This follows from the finding that the [Fe/H]$<-1$ halo within a few kpc from the Sun is dominated by GSE \citep[e.g.,][]{DiMatteo19,Bonaca20,An21}. Further, not just GSE, but almost all the disrupted dwarfs appear systematically $\alpha$-enhanced compared to their surviving peers. We make a more controlled comparison at fixed stellar mass that makes this apparent in the following figures. 

In Figure \ref{fig:mzr} we depict stellar masses and stellar metallicities. As a gray dashed line we show the $z=0$ MZR estimated from 35 MW and local group dwarfs, most of which are not shown in Figure \ref{fig:mzr} \citep{Kirby13}. The disrupted dwarfs fall along a fairly tight sequence that runs almost parallel to the $z=0$ MZR but is offset to lower metallicities. The median offset, computed by subtracting the expected [Fe/H] as per the $z=0$ MZR from the observed [Fe/H], is $0.27^{+0.05}_{-0.02}$ dex. Formally, we fit the disrupted dwarf MZR to be:
\begin{align}
\label{eqn:mzr}
\rm{[Fe/H]} = -2.11^{+0.11}_{-0.12} + 0.36^{+0.12}_{-0.04}\log{\left(\frac{\textit{M}_{\star}}{10^{6} \textit{M}_{\rm{\odot}}}\right)}.
\end{align}

The slope for the disrupted dwarf MZR is almost identical to the $z=0$ \citet[][]{Kirby13} MZR for surviving dwarfs ($0.30^{+0.02}_{-0.02}$), and the normalization is offset to lower metallicities by $\approx0.4$ dex ($-2.11^{+0.11}_{-0.12}$ vs. $-1.69^{+0.04}_{-0.04}$). Sgr, which straddles the boundary between surviving and disrupted, as it is still in the process of being stripped and has been forming stars as recently as 2 Gyrs ago \citep[e.g.,][]{Alfaro-Cuello19} as well as I'itoi, which was likely accreted significantly earlier than the rest of our sample are clear outliers in Fig. \ref{fig:mzr} and excluded from this fit. Including them results in a steeper slope of $0.50^{+0.08}_{-0.08}$ and a similar offset of $-2.30^{+0.15}_{-0.12}$.

In the right panel of Figure \ref{fig:mzr} we plot [$\alpha$/Fe] vs. stellar mass for the dwarf samples. At fixed stellar mass, almost all the disrupted dwarfs are systematically $\alpha$-enhanced, forming a sequence that lies above their surviving peers. That is, not only is the stellar halo $\alpha$-enhanced compared to the surviving dwarfs in aggregate, but this trend also holds when the halo is resolved into distinct galaxies. Not just GSE, but all the disrupted dwarfs in the inner halo are systematically $\alpha$-enhanced compared to the surviving dwarfs. UMin and Orphan/Chenab are instructive exceptions and are discussed in \S\ref{sec:whymetalpoor}.

\begin{figure}[t]
\centering
\includegraphics[width=\linewidth]{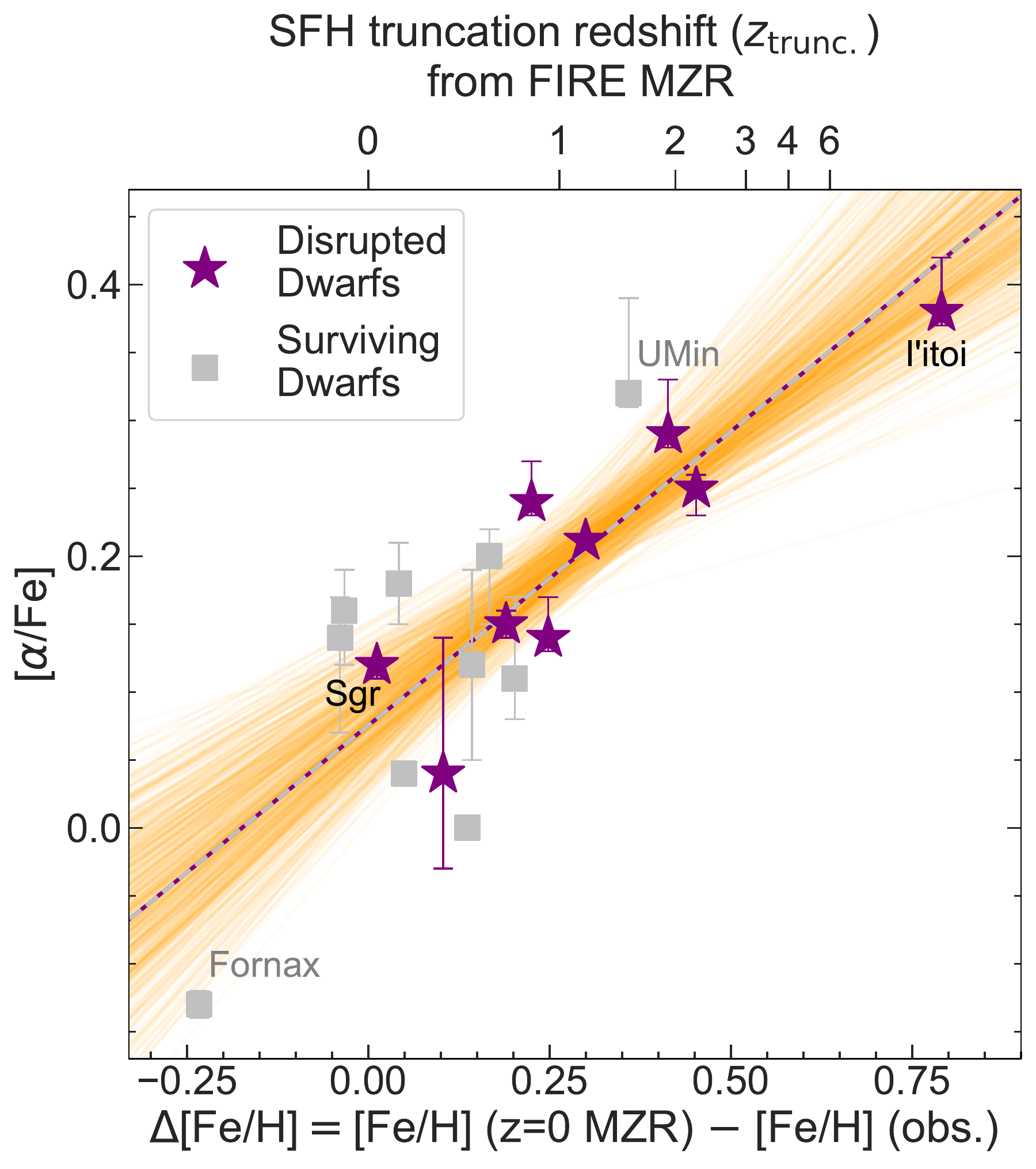}
\caption{[$\alpha$/Fe] vs. $\Delta$[Fe/H], where the latter measures how metal-poor a galaxy is compared to the expectation from the $z=0$ MZR (see Eqn. \ref{eqn:deltafeh}). The dashed line is fit to both surviving and disrupted dwarfs. Fits to 100 bootstrap samples of the data are shown as transparent orange lines. The more metal-poor a dwarf is compared to the expectation from the the $z=0$ MZR (i.e., a higher $\Delta$[Fe/H]), the more $\alpha$-enhanced it is likely to be. In hydrodynamical simulations $\Delta$[Fe/H] correlates with the redshift of star-formation truncation ($z_{\rm{trunc}}$), and so the above plot may be interpreted as a trend of higher [$\alpha$/Fe] on average at higher redshifts. Shown at the top are $z_{\rm{trunc}}$ values corresponding to $\Delta$[Fe/H] from the FIRE simulations (Eqn. \ref{eqn:afe}, \citealt[][]{Ma16MZR}). We highlight Sgr and I'itoi as extremes among the disrupted dwarfs, and UMin and Fornax as extremes among the surviving dwarfs. While Fornax and Sgr have formed stars till very recently ($z\approx0$), UMin and I'itoi likely assembled their mass rapidly and very early.}
\label{fig:delta}
\end{figure}

In Figure \ref{fig:delta} we plot [$\alpha$/Fe] as a function of $\Delta$[Fe/H], i.e., how metal-poor a galaxy is compared to the expectation from the $z=0$ \citet[][]{Kirby13} MZR for its stellar mass. The motivation for introducing $\Delta$[Fe/H] comes from high-$z$ observations \citep[e.g.,][]{Henry21,Topping21,Sanders21} and hydrodynamical simulations \citep[e.g.,][]{Ma16MZR,Dave17,deRossi17,Torrey19,Langan20} that show that the MZR evolves along parallel tracks towards lower metallicities at higher redshifts (discussed further in \S\ref{sec:highzarchaeology}). We interpret $\Delta$[Fe/H] as tracking the redshift at which the galaxy falls on the MZR, i.e., a higher $\Delta$[Fe/H] implies the galaxy would be a typical galaxy at a higher redshift. We compute $\Delta$ [Fe/H] as follows:

\begin{align}
\label{eqn:deltafeh}
    \Delta \rm{[Fe/H]} &= \rm{[Fe/H]} \ \rm{(MZR\ |\ \it{M}_{\rm{\star}}\ \rm{obs.})} - \rm{[Fe/H]\ (obs.)} \nonumber\\
    &= -1.69 + 0.30 \log{\left(\frac{M_{\star} \rm{(obs.)}}{10^{6} M_{\rm{\odot}}}\right)} - \rm{[Fe/H] (obs.)},
\end{align}
where [Fe/H] (obs.) and $M_{\star}$ (obs.) are the observed metallicity and stellar mass, as listed in Table \ref{table:summary}. To translate $\Delta$[Fe/H] into the redshift when the dwarf was located on the MZR ($z_{\rm{trunc}}$, i.e., the redshift of star-formation truncation), we adopt the MZR evolution from the FIRE simulations \citep[][]{Ma16MZR} that is consistent with latest constraints for $M_{\rm{\star}}\gtrsim10^{9} M_{\rm{\odot}}$ galaxies (reviewed in \citealt[][]{Maiolino19}):

\begin{align}
\label{eqn:ztrunc}
    z_{\rm{trunc}}=-2\ln\left(1 - \Delta\rm{[Fe/H]}/0.67\right).
\end{align}

This conversion allows us to associate each dwarf with a redshift when it likely finished assembling its mass. The key caveat for this conversion is that across $z\approx1-3$ a modest amount of intrinsic scatter is observed in the MZR at $M_{\rm{\star}}>10^{9}M_{\rm{\odot}}$ ($\approx0.1$ dex, \citealt[][]{Sanders21}), but the magnitude of scatter in lower mass galaxies is currently unconstrained. As a confidence-inspiring cross-check, for the two disrupted dwarfs with well-measured star-formation histories (SFHs), GSE and Sgr, we find $z_{\rm{trunc}}=1.2$ and $z_{\rm{trunc}}<0.1$, in excellent agreement with their reported SFHs \citep[e.g.,][]{Bonaca20,Alfaro-Cuello19}.

In Figure \ref{fig:delta} we see $\Delta$[Fe/H] correlates remarkably well with [$\alpha$/Fe] for both disrupted and surviving dwarfs. Systems with large $\Delta$[Fe/H], that likely assembled their stars at higher redshifts are more $\alpha$-enhanced. The relationship, fit to surviving as well as disrupted dwarfs is described as follows:

\begin{align}
\label{eqn:afe}
    [\alpha/\rm{Fe}]= 0.43^{+0.09}_{-0.09}\times\Delta\rm{[Fe/H]} + 0.08^{+0.03}_{-0.03}.
\end{align}

\begin{figure}[t]
\centering
\includegraphics[width=\linewidth]{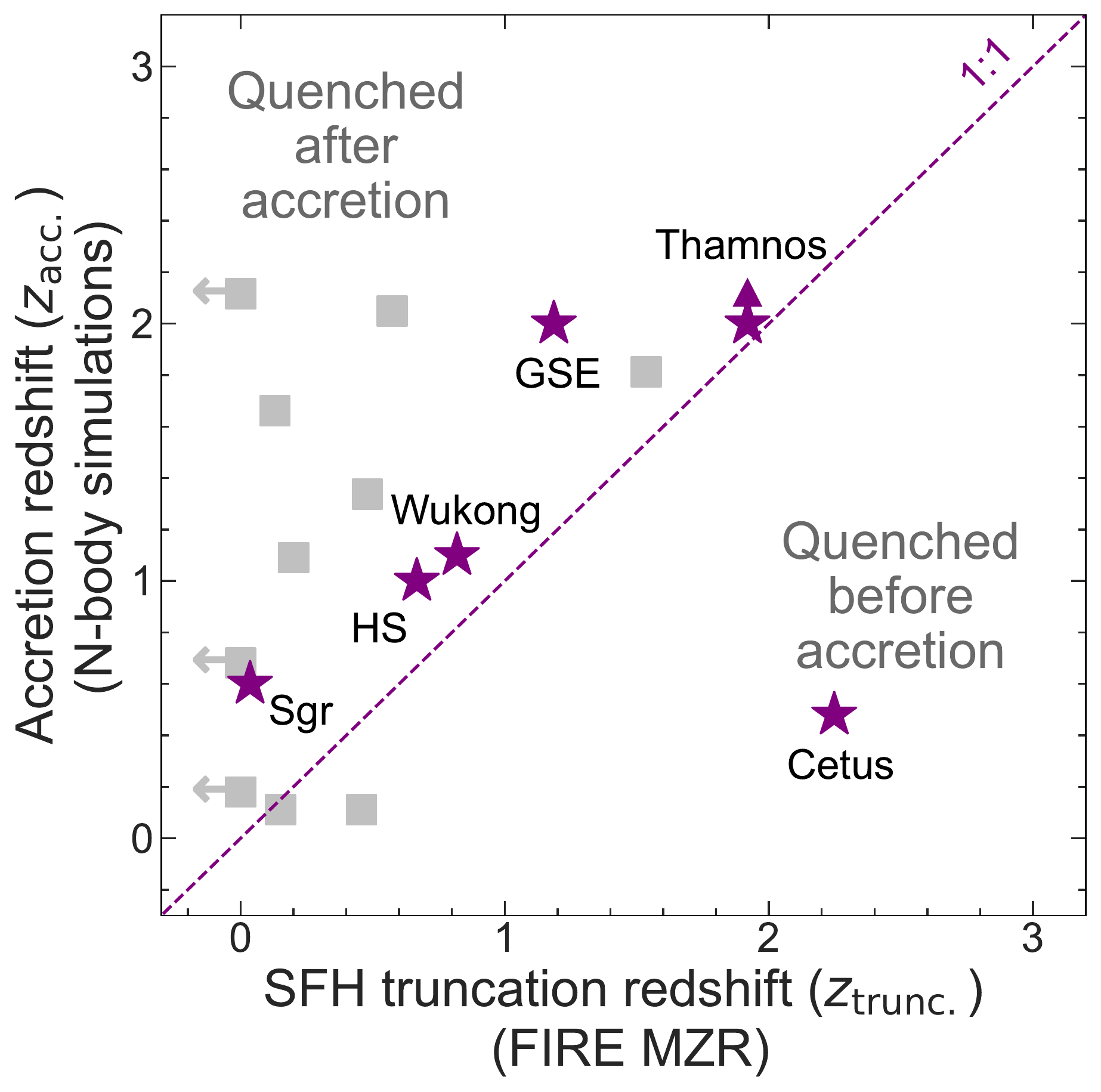}
\caption{Accretion redshift ($z_{\rm{acc}}$) from N-body simulations vs. the star-formation truncation redshift ($z_{\rm{trunc}}$) for disrupted (purple stars) and surviving (gray squares) dwarfs. The surviving dwarfs with $z_{\rm{trunc}}$ consistent with recent/ongoing star-formation (i.e., $\Delta$[Fe/H]$>0$) are shown with gray arrows. The correlation between these quantities for the disrupted dwarfs is heartening given that they are estimated independently -- $z_{\rm{acc}}$ from dynamics and $z_{\rm{trunc}}$ from chemistry. The $1:1$ line is shown in purple. The clustering of purple points along this line argues for the truncation of SFH in disrupted dwarfs being due to their interactions with the Milky Way. No sources, except for Cetus, are consistent with being quenched significantly before accretion.}
\label{fig:sims}
\end{figure}

In Figure \ref{fig:sims} we compare the star-formation truncation redshift inferred via chemistry ($z_{\rm{trunc}}$, Eqn. \ref{eqn:ztrunc}) and the accretion redshift inferred via dynamics ($z_{\rm{acc}}$). For the disrupted dwarfs we adopt $z_{\rm{acc}}$ (i.e., when the first pericentric passage occurs) from the simulations listed in \S\ref{sec:data}, whereas for the surviving dwarfs we rely on \citet[][]{Fillingham19} who inferred $z_{\rm{acc}}$ by comparing \textit{Gaia} satellite dynamics \citep[][]{Fritz18} to the Phat ELVIS simulations of MW-mass galaxies \citep[][]{Kelley19}. Sgr is an exception, for which we adopt $z_{\rm{acc}}$ from \citealt[][]{Lian20,Ruiz-Lara20} who measured starbursts in the MW disk that are likely coincident with Sgr's pericentric passages \citep[][]{dicintio21}. For Thamnos we interpret the arguments in \citet[][]{Koppelman19} as implying $z_{\rm{acc}}>2$. In particular, for a $\approx10^{6}M_{\rm{\odot}}$ galaxy to have sunk as deep as Thmanos in the potential -- $E_{\rm{tot}}\approx-1.35\times10^{5}$ km$^{2}$ s$^{-2}$, the lowest orbital energy of all the galaxies considered here -- strongly implies $z_{\rm{acc}}>2$ (see e.g., first row of Fig. 3 in \citealt[][]{Pfeffer20}). 

Given the entirely independent methods that go into deriving $z_{\rm{acc}}$ and $z_{\rm{trunc}}$, it is encouraging that both the disrupted and surviving dwarfs are not randomly scattered in Figure \ref{fig:sims}. Further, for the disrupted dwarfs $z_{\rm{trunc}}\propto z_{\rm{acc}}$, consistent with these systems being quenched shortly after accretion. Cetus is the exception, with $z_{\rm{trunc}}>>z_{\rm{acc}}$, which may be linked to uncertainty in its derived stellar mass (\S\ref{sec:datadisrupted}). There are caveats. Systematic uncertainties are significant and poorly understood -- e.g., the simulations informing $z_{\rm{acc}}$ do not account for an evolving MW potential with distortions produced by the Magellanic Clouds \citep[e.g.,][]{Garavito-Camargo19,Garavito-Camargo21,Cunningham20,Conroy21} that likely biases the inferred $z_{\rm{acc}}$ in a non-trivial fashion. On the other axis, $z_{\rm{trunc}}$ is inferred from one suite of hydrodynamical simulations (FIRE, \citealt[][]{Ma16MZR}), though we have verified excellent agreement for Sgr and GSE.

\section{Discussion}
\label{sec:discussion}

\subsection{Star-formation timescales explain why the disrupted dwarfs are metal-poor and $\alpha$-enhanced}
\label{sec:whymetalpoor}

At fixed mass, the disrupted dwarfs are systematically metal-poor compared to the $z=0$ MZR and $\alpha$-enhanced compared to the surviving dwarfs (Figure \ref{fig:mzr}). This difference may be understood in terms of star-formation timescales -- to grow to a given stellar mass, the disrupted dwarfs, with typical $z_{\rm{trunc}}\approx1-2$, formed stars on a shorter timescale (i.e., pre-disruption) compared to dwarfs of similar mass that survived to $z=0$ (e.g., GSE compared to SMC/LMC/Sgr, Sequoia/Wukong compared to Fornax). The compressed star-formation history leaves less time for [$\alpha$/Fe] elevated by Type II supernovae to be diluted by Fe from Type I supernovae \citep[e.g.,][]{Tinsley79, Robertson05,Font06}. Let us consider a few informative outliers that add to this picture.

UMin is the most metal-poor and $\alpha$-enhanced of the surviving dwarfs, and lies closer to the disrupted dwarfs in all our figures (e.g., Fig. \ref{fig:delta}). Of the surviving dwarfs, it has among the most rapid star-formation timescales, assembling almost all its mass in $\approx400$ Myrs \citep[e.g.,][]{Kirby11aFe,Weisz14sfh,Gallart15}. UMin was likely accreted at $z\approx2$ \citep[][]{Fillingham19} and quenched shortly after by e.g., ram pressure stripping. Its circular orbit ($e\approx0.4-0.5$, \citealt[][]{Fritz18,Li21,Battaglia22}) and low mass relative to the other disrupted dwarfs may have kept it from being disrupted, since these properties prolong the dynamical friction timescale \citep[e.g.,][]{Amorisco17,Naidu21, Vasiliev22}. We also make note of Crater 2 and Antlia 2, which lie on the disrupted dwarf MZR and have the highest $\Delta$[Fe/H] (0.34, 0.28) after UMin (0.36). These extended, low surface brightness dwarfs do show signs of disruption \citep[e.g.,][]{Ji21} and may also have been quenched early -- our empirical relation predicts an elevated [$\alpha$/Fe] in these galaxies.

Conversely, Sgr and Orphan/Chenab have the lowest $\Delta$[Fe/H], [$\alpha$/Fe], and $z_{\rm{trunc}}$ of all the disrupted dwarfs, in line with them being relatively recent arrivals to the MW with extended star-formation histories. This is consistent with \citet[][]{Panithanpaisal21}, who analysed the FIRE2 simulations \citep[][]{Wetzel16,Garrison-Kimmel19} to conclude that dwarf galaxies still observable as cogent stellar streams are often chemically similar to the surviving dwarfs. Note that Orphan/Chenab has the highest orbital energy (second only to Sgr) of all systems analyzed in this work -- recently infalling low-$\alpha$ streams like Orphan/Chenab may comprise a large fraction of the halo beyond 50 kpc that we do not analyze in this work (in fact, \citealt[][]{Font06} suggest low [$\alpha$/Fe] as a search strategy for such dwarfs).

These examples highlight that despite the broad differences we find in this work between the surviving and the disrupted dwarfs, there is overlap in their stellar populations -- particularly when considering the disrupted dwarfs at $>50$ kpc not in our sample, which likely resemble the surviving dwarfs. While the star-formation timescale is informative, it may not be the entire story -- e.g., Sculptor/Sextans appear to also have rapidly built up their mass at comparable redshifts to the disrupted dwarfs \citep[e.g.,][]{Kirby11aFe,Weisz14sfh,Gallart15}, but are not as $\alpha$-enhanced as I'itoi/Thamnos, though the difference is not as stark as at $>10^{7} M_{\rm{\odot}}$. Further, a deeper question persists -- \textit{why} were galaxies like GSE/Wukong/Thamnos able to form as much mass as SMC/Fornax/Leo I in a fraction of the time? We tackle these issues in the following section.

\subsection{Different formation channels for disrupted and surviving dwarfs?}
\label{sec:butwhymetalpoor}

A generic prediction of $\Lambda$CDM simulations is that disrupted dark matter subhalos are accreted and destroyed at higher redshifts than surviving subhalos \citep[e.g.,][]{Gao04, Zentner05,vandenBosch05}. This trend is also borne out in zoom-in simulations of MW-like stellar halos that find disrupted dwarfs have higher accretion redshifts (typical $z_{\rm{acc}}>1$) compared to their surviving peers \citep[e.g.,][]{Fattahi20}. Subhalos that assembled their mass at higher redshifts are more concentrated with a higher characteristic overdensity ($\propto(1+z)^{3}$, \citealt[][]{MvdBW}), and are also likely to inhabit clustered, overdense environments (see \citealt[][]{Wechsler18} for a recent review).

In light of this, we offer the following scenario. The disrupted dwarfs were likely born in concentrated dark matter halos in a dense, gas-rich environment close to the MW. Their halos resisted the loss of enriched gas to feedback. They were able to sustain high star-formation efficiencies for large fractions of their lifetimes, resulting in high [$\alpha$/Fe]. Nonetheless, their proximity to the MW resulted in prompt accretion within a few Gyrs after formation, which is reflected in their high $z_{\rm{trunc}}$ and $z_{\rm{acc}}$ (e.g., GSE, Thamnos, I'itoi). On the other hand, most surviving dwarfs, as well as disrupted dwarfs accreted recently (e.g., Orphan/Chenab) began their journey towards our galaxy from farther low-density regions and/or were born in less-concentrated halos that found it difficult to retain enriched gas (explaining differences between I'itoi/Thamnos vs. Sculptor/Sextans). The longer journeys and less concentrated halos directly translate to extended star-formation histories, low [$\alpha$/Fe], and lower $z_{\rm{acc}}$. We note that \citet[][]{Gallart15} made an analogous argument based on star-formation histories and dynamics to explain why ``fast"-forming dwarfs like UMin differed from ``slow"-forming dwarfs like Fornax.


\subsection{Timing accretion, quenching, and disruption}
\label{sec:sequencing}
All surviving and disrupted dwarfs (Cetus being the one significant exception) have $z_{\rm{acc.}} > z_{\rm{trunc.}}$ -- that is, these dwarfs were first accreted onto the Milky Way, and only subsequently quenched with some delay. Note that $z_{\rm{trunc}}\geq z_{\rm{acc}}$ is also in principle possible due to e.g., self-quenching via feedback or group preprocessing prior to infall \citep[e.g.,][]{Panithanpaisal21,Samuel22} -- Figure \ref{fig:sims} suggests this must be rare at $M_{\star}\gtrsim10^{6} M_{\rm{\odot}}$. 

The clustering of the disrupted dwarfs close to the $1:1$ line in Fig. \ref{fig:sims} implies that whatever the mechanisms for quenching after infall (e.g., ram pressure stripping by the circumgalactic medium, tidal dissolution), they allow time for a few pericenters and some additional star-formation. This additional star-formation is predicted to be only $\approx10\%$ of the mass, \citealt[][]{Fattahi20}, and perhaps seen empirically for GSE between 8-10 Gyrs in \citealt[][]{Bonaca20}. This chronology is reminiscent of the \citet[][]{Wetzel13} ``delayed-then-rapid" quenching scenario for satellites falling into galaxy clusters whose star-formation is abruptly truncated, but only a handful of pericenters after infall -- in our case the MW plays the role of the cluster. Interestingly, the difference between $z_{\rm{acc}}$ and $z_{\rm{trunc}}$ is larger, and shows more scatter for the surviving dwarfs -- this may reflect a combination of their orbital and structural parameters that have managed to keep dynamical friction at bay and delayed quenching.

That $z_{\rm{acc}}>z_{\rm{trunc}}$ for virtually all sources in Figure \ref{fig:sims} is an archaeological way of inferring that $\gtrsim10^{6}M_{\rm{\odot}}$ field dwarfs are rarely quenched, and that quenching is intimately linked to interactions with a massive host. While the fraction of field dwarfs that are observed to be quenched at $z\approx0$ is very low \citep[e.g.,][]{Geha12}, our results suggest that this continues to be the case even at higher redshifts.

The timing constraints from this work also bracket the epoch when the stellar halo between $\approx10-50$ kpc was built up. The disrupted dwarfs in our sample largely arrived in the Galaxy at $z\gtrsim1$, with the $z_{\rm{trunc.}}$ of any individual source typically acting as a lower limit on the accretion redshift (Fig. \ref{fig:sims}). This is in excellent agreement with simulations that predict the majority of the disrupted dwarfs have $z_{\rm{acc}}>1$ \citep[e.g.,][]{Fattahi20}.

\subsection{High-$z$ stellar abundances: a Galactic archaeology approach}
\label{sec:MZRevol}
\label{sec:highzarchaeology}

The current redshift frontier for direct, in-situ studies of stellar and gas-phase abundances is $z\approx2-3$\footnote{A handful of constraints at higher redshifts ($z\approx4-8$) exist on individual objects \citep[e.g.,][]{Shapley17,Matthee21b} or via calibrations where the current systematic uncertainties are likely on the order of the expected evolution in the abundance trends \citep[e.g.,][]{Faisst16c,Jones20}. This is set to change soon thanks to \textit{JWST} \citep[][]{Shapley21,Strom21,Curti21}.} \citep[e.g.,][]{Steidel16,Strom18,Cullen19,Shapley19,Topping20,Sanders20,Sanders21,Cullen21,Kashino22,Strom22}. A key trend reported in these studies is that galaxies at these redshifts are significantly metal-poor compared to their $z=0$ counterparts. The MZR is offset to lower metallicities, but runs almost parallel to the $z=0$ MZR. Further, these studies infer $\alpha$-enhancement at $z\approx2-3$ on the order of [$\alpha$/Fe]$
\approx0.4$, i.e., ($\alpha$/Fe)$\approx2.5\times$($\alpha$/Fe)$_{\odot}$, based on e.g., comparisons of rest-UV spectra with stellar population synthesis models. 

There are two main limitations to these findings, where MW halo constraints are uniquely complementary. Due to observational feasibility, the stellar mass range probed at $z\approx2-3$ is $\gtrsim10^{9} M_{\rm{\odot}}$. Whether the reported trends extend to lower mass systems is a key open question with wide-ranging implications e.g., for reionization vis-a-vis the expected ionizing power of low luminosity systems that dominate integrals of UV luminosity functions (e.g., \citealt[][]{Finkelstein19, Naidu20a}; \citetalias{Paper2} \citeyear{Paper2}). Further, calibrations of strong-line indicators required to convert observed nebular emission to abundances are challenging \citep[e.g.,][]{Sanders21}, and several critical unknowns in stellar models of massive, metal-poor stars remain (see \citealt[][]{Eldridge22} for a recent review) for which observational constraints are notoriously sparse (only a handful of O stars at [Fe/H]$\lesssim-1$ have UV spectra, \citealt[][]{Telford21}). Given these systematics, any independent lines of evidence would inspire greater confidence.

Dynamical constraints (N-body simulations) suggest typical accretion and disruption redshifts of $z\approx1-2$ for the disrupted dwarfs in our sample (Figure \ref{fig:sims}). From Figure \ref{fig:mzr} and our fit in Eqn. \ref{eqn:mzr} we see that the disrupted dwarfs support the trend of an MZR with a similar slope, but an offset to lower metallicities continuing into the $10^{6}-10^{9} M_{\rm{\odot}}$ range at these redshifts. The magnitude of the offset ($\approx$0.3-0.4 dex, \S\ref{sec:results}) is slightly lower than the 0.6 dex reported at $z\approx3.5$ \citep[][]{Cullen19,Cullen21} and 0.8 dex reported at $z\approx2.2$ \citep[][]{Kashino22}, which is not surprising given that we are probing slightly lower redshifts.

As for [$\alpha$/Fe], we can use our fit in Figure \ref{fig:delta} (Eqn. \ref{eqn:afe}) to predict the expected $\alpha$-enhancement at $z\approx2-3$. For an average $\Delta$[Fe/H] of $0.6-0.8$ \citep[][]{Cullen19,Kashino22}, we predict [$\alpha$/Fe]$\approx0.35-0.45$. This value is in remarkable agreement with the inferred [$\alpha$/Fe] for $\approx10^{9}-10^{11} M_{\rm{\odot}}$ star-forming galaxies \citep[e.g.,][]{Cullen21,Strom22,Kashino22}, and is an entirely independent line of evidence for elevated [$\alpha$/Fe] of this magnitude, and in low-mass galaxies, at $z\approx2-3$. The similar mapping between $\Delta$[Fe/H] and [$\alpha$/Fe] across such a wide-range in stellar mass ($\approx10^{6} -10^{11}\, M_{\rm{\odot}}$) argues for weak mass-dependence in these abundance trends.

\section{Summary \& Outlook}
This paper presented a comparative analysis of the MW's surviving and disrupted dwarfs. While the surviving dwarfs are relatively well-characterized, the majority of the disrupted dwarfs have only recently been discovered thanks to \textit{Gaia} and spectroscopic surveys like H3. We derive and compile stellar masses and abundances for almost all known disrupted and surviving dwarfs with $M_{\rm{\star}}\approx10^{6}-10^{9} M_{\rm{\odot}}$ [\S\ref{sec:data}, Table \ref{table:summary}, \ref{table:surviving}]. We find the following:

\begin{itemize}
    \item At fixed stellar mass, the disrupted dwarfs are relatively metal-poor and $\alpha$-enhanced compared to the surviving dwarfs. [Fig. \ref{fig:mzr}]
    
    \item The disrupted dwarfs define a mass-metallicity relation that runs parallel to the $z=0$ MZR, but is offset towards lower metallicities by $\approx0.3-0.4$ dex. [Fig. \ref{fig:mzr}, Eqn. \ref{eqn:mzr}]
    
    \item Both surviving and disrupted dwarfs follow a $\Delta$[Fe/H] - [$\alpha$/Fe] relation -- the more metal-poor a galaxy is compared to the $z=0$ MZR, the more $\alpha$-enhanced it is. [Fig. \ref{fig:delta}, Eqn. \ref{eqn:afe}].
    
    \item Comparison of chemically inferred star-formation truncation redshifts ($z_{\rm{trunc}}$) with dynamically inferred accretion redshifts ($z_{\rm{acc}}$) shows that almost all dwarfs are quenched only after being accreted ($z_{\rm{acc}}>z_{\rm{trunc}}$). The disrupted dwarfs are quenched promptly within a few pericenters after accretion ($z_{\rm{acc}}\approx z_{\rm{trunc}}$), whereas the surviving dwarfs display longer lag times. This is an archaeological way of inferring that field dwarfs are rarely quenched, not just at $z\approx0$ as has been established by direct observations, but also at higher redshifts. [Fig. \ref{fig:sims}, \S\ref{sec:sequencing}]
    
    \item The chemical differences between the surviving and disrupted dwarfs are likely a consequence of their star-formation efficiencies, with the disrupted dwarfs having built up their mass relatively rapidly (typical $z_{\rm{trunc}}\approx1-2$). We propose this is because the dark matter halos hosting the disrupted dwarfs formed at higher redshifts, were more concentrated, and were embedded in dense environments proximal to the Milky Way -- all conditions conducive to high star-formation efficiencies. [\S\ref{sec:whymetalpoor}, \S\ref{sec:butwhymetalpoor}]
    
    \item We place archaeological constraints on high-$z$ stellar chemistry in a mass regime currently beyond the reach of direct studies. The disrupted dwarfs with typical accretion and disruption redshifts of $z\approx1-2$ have $\Delta$[Fe/H]$\approx0.3-0.4$, supporting extrapolations of the MZR evolution at $M_{\rm{\star}}\gtrsim10^{9} M_{\rm{\odot}}$ to lower masses. Our $\Delta$[Fe/H]-[$\alpha$/Fe] relation (Eqn. \ref{eqn:afe}) predicts $z\approx2-3$ galaxies have [$\alpha$/Fe]$\approx0.4$, in remarkable agreement with recently inferred values for $M_{\rm{\star}}>10^{9} M_{\rm{\odot}}$ systems. Our findings suggest these abundance trends are mass-independent across $M_{\rm{\star}}\approx10^{6}-10^{11} M_{\rm{\odot}}$. [\S\ref{sec:highzarchaeology}]
\end{itemize}

We emphasize once again that our findings apply to the disrupted dwarfs catalogued by the H3 Survey in \citealt[][]{Naidu20} ($r_{\rm{gal}}\approx6-50$ kpc, $|b|>40^{\rm{\circ}}$) -- disrupted dwarfs lying entirely at larger distances were likely accreted recently ($z\lesssim0.5$) and resemble the surviving dwarfs, whereas those entirely buried in the inner galaxy were preferentially accreted early ($z\gtrsim2$), and are likely even more metal-poor and $\alpha$-enhanced than than the dwarfs studied here. Some of the arguments introduced (e.g., translating $\Delta$[Fe/H] to [$\alpha$/Fe], the archaeological connection to direct high-$z$ studies) are promising, but nonetheless preliminary, since they are based on a limited sample. However, this is set to change imminently. For every $\gtrsim10^{5} M_{\rm{\odot}}$ surviving satellite the Milky Way accreted, there were likely two that were disrupted \citep[e.g.,][]{Fattahi20}. In other words, there may be $\gtrsim100$ disrupted dwarfs dissolved across the Galaxy awaiting discovery. While our current sample is preferentially sampling the most massive of these sources, we are on the brink of a major expansion. In the coming decade there is a clear path to much larger halo samples than the H3 Survey has amassed thanks to the DESI \citep[][]{Prieto20}, SDSS-V \citep[][]{SDSSV}, 4MOST \citep[][]{4MOSTlowres}, and WEAVE \citep[][]{Dalton12} surveys, which promise to complete our census of dwarf galaxies, both surviving and disrupted, out to the virial radius of the Galaxy.

\facilities{MMT (Hectochelle), \textit{Gaia}}

\software{
    \package{IPython} \citep{ipython},
    \package{matplotlib} \citep{matplotlib},
    \package{numpy} \citep{numpy},
    \package{scipy} \citep{scipy},
    \package{jupyter} \citep{jupyter},
    \package{gala} \citep{gala1, gala2},
    \package{Astropy}
    \citep{astropy1, astropy2},
    }
    
\acknowledgments{RPN gratefully acknowledges an Ashford Fellowship granted by Harvard University. CC and PC acknowledge
support from NSF grant NSF AST-2107253. YST acknowledges financial support from the Australian Research Council through DECRA Fellowship DE220101520. NC acknowledges support from NSF AST-1812461. We thank the Hectochelle operators and the CfA and U. Arizona TACs for their continued support of the H3 Survey. 

This paper uses data products produced by the OIR Telescope Data Center, supported by the Smithsonian Astrophysical Observatory. The computations in this paper were run on the FASRC Cannon cluster supported by the FAS Division of Science Research Computing Group at Harvard University. This work has made use of data from the European Space Agency (ESA) mission
{\it Gaia} (\url{https://www.cosmos.esa.int/gaia}), processed by the {\it Gaia}
Data Processing and Analysis Consortium (DPAC,
\url{https://www.cosmos.esa.int/web/gaia/dpac/consortium}) \citep{dr2ack1, dr2ack2}. Funding for the DPAC
has been provided by national institutions, in particular the institutions
participating in the {\it Gaia} Multilateral Agreement.}

\bibliography{MasterBiblio}
\bibliographystyle{apj}

\end{CJK*}
\end{document}